\documentclass[prl,aps,preprint]{revtex4}
\usepackage{graphicx}
\usepackage{latexsym}
\usepackage{amsmath}
\usepackage{epstopdf}
\begin{document}

\title{Anyons from fermions with conventional two-body interactions}

\author{Yue Yu  }
\affiliation{Institute of Theoretical Physics, Chinese Academy of
Sciences, P.O. Box 2735, Beijing 100190, China}
\author{ Yi Li}
\affiliation{Department of Physics, Fudan University, Shanghai
200433, China}

\begin{abstract}
{ Emergent anyons are the key elements of the
 topological quantum computation and topological
quantum memory. We study a two-component fermion model with
conventional two-body interaction in an open boundary condition
and show that several subsets in the low-lying excitations obey
the same fusion rules as those of the toric code model. Those
string-like non-local excitations in a given subset obey mutual
semionic statistics. We show how to peel off one of such subset
from other degenerate subsets and manipulate anyons in cold
dipolar Fermi atoms or cold dipolar fermionic heteronuclear
molecules in optical lattices by means of the established
techniques.}
\end{abstract}

\date{today}


 \maketitle

\noindent{\it Introductions} Anyons are long wistful objects in
two-dimensional condensed matter systems \cite{anyon,wil}.
Interesting to explore anyons is currently renewed because of the
potential application of anyons in topological quantum computation
and topological quantum memory \cite{kitaev,rev}. The researches
mainly focus on two topics: non-abelian fractional quantum Hall
states \cite{mr,rr,free} and particular lattice spin models, e.g.,
Kitaev toric code model \cite{kitaev}, Levin-Wen model \cite{lw}
and Kitaev honeycomb-lattice spin model \cite{kitaev1}.

Experimentally exciting, manipulating and detecting abelian anyons
have been suggested or tried for the toric code model
\cite{han,pan} and for the insulating phase of Kitaev
honeycomb-lattice model \cite{zd}. Although these non-trivial
tries are interesting, the reliable evidence for the existence of
anyons still lacks since neither the toric code model nor
honeycomb-lattice spin model is easy to be realized owing to those
unconventional interactions between constitution particles, e.g.,
cold atoms \cite{duan}.

\begin{figure}[htb]
\includegraphics[width=2.5cm]{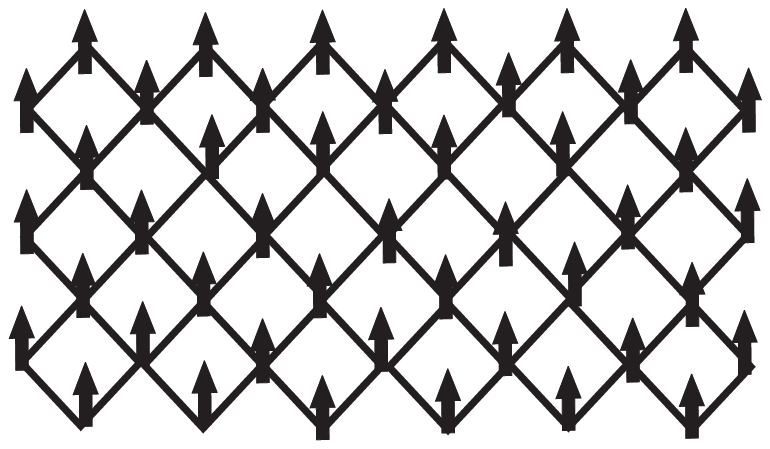}
\includegraphics[width=2.5cm]{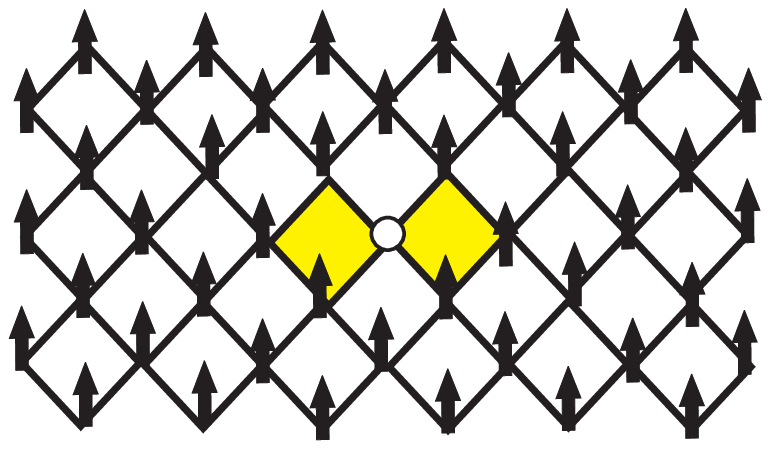}
\includegraphics[width=2.5cm]{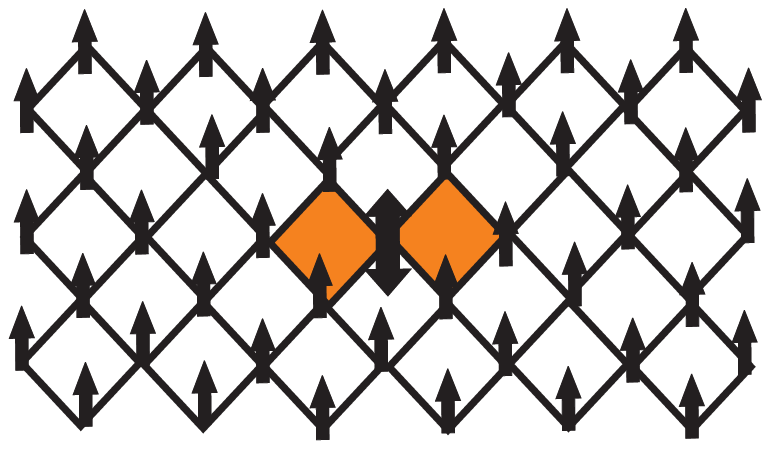}
\includegraphics[width=2.5cm]{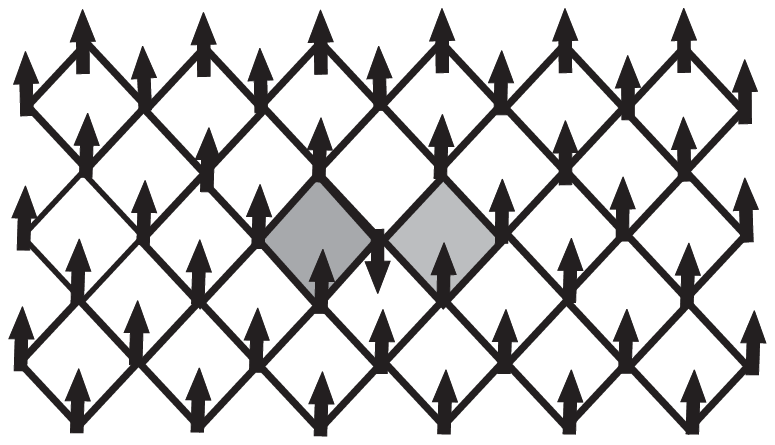}
\includegraphics[width=2.5cm]{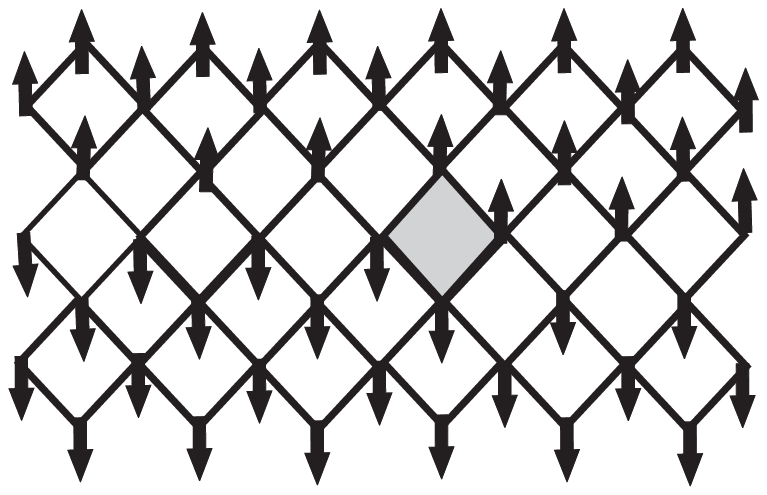}
\includegraphics[width=2.5cm]{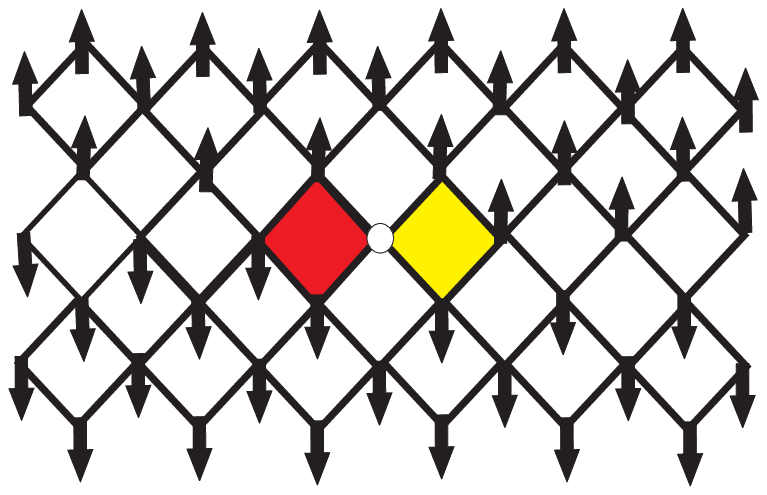}
\includegraphics[width=2.5cm]{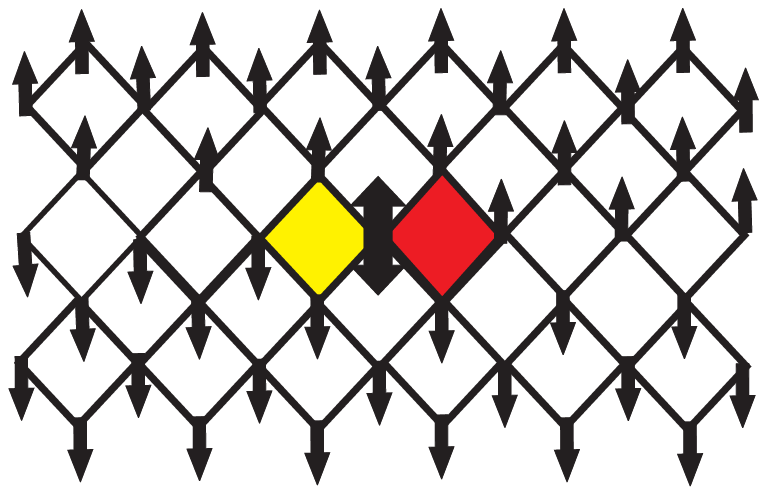}
\includegraphics[width=2.5cm]{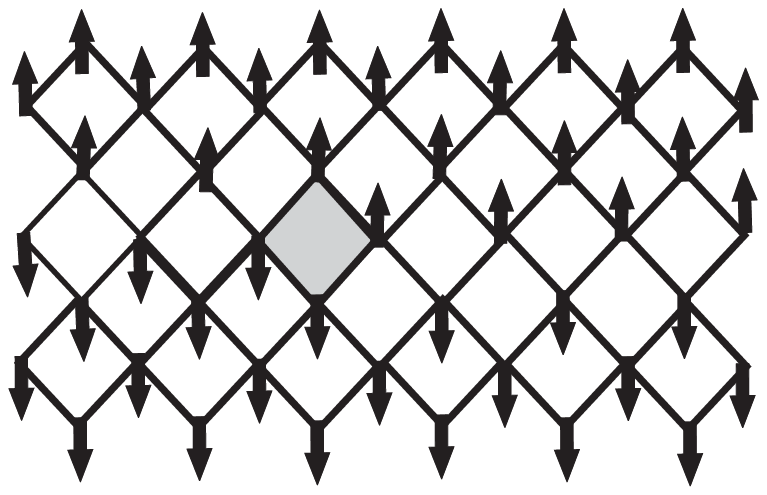}
 \caption{\label{fig:Fig.1} (Color on line)
The ground state ( taking $|G_\uparrow\rangle$ as an example), and
the low-lying excitations in a set of decoupled Ising spin chains
which form a square lattice. From left to right and up to down,
they are $|G_\uparrow\rangle,|{\cal H}_{i_a}\rangle,|{\cal
D}_{i_a}\rangle,|{\cal F}_{i_a}\rangle,|{\cal W}_P\rangle,|{\cal
W}^h_{P,P'}\rangle,|{\cal W}^d_{P,P'}\rangle$ and $|{\cal
W}_{P'}\rangle$.  Up- and down-arrows label the fermion with
spin-up and spin down. Empty circle is unoccupied site and up-down
arrow is double occupied. The white plaquette $P"$ has
$(G^{s}_{P"},G^{\bar s}_{P"})=(1,1)$, the yellow has $(-1,1)$, the
red has $(1,-1)$, and the grey has $(-1,-1)$.}
\end{figure}

A widely interesting question is: do we have a simple system with
conventional two-body interactions whose non-local elementary
excitations are anyons? We here study a two-component fermion
system which is a set of decoupled Ising chains\cite{yuli}. These
chains form a two-dimensional square lattice with each chain along
the horizontal diagonal direction (See Fig. \ref{fig:Fig.1}). The
fermions within a chain interact with on-site repulsive and
nearest neighbor attractive potentials. Nussinov and Ortiz
\cite{no} have found the decoupled Ising chains are of the same
spectrum as the toric code model and Liven-Wen model . However,
the topological order is trivial when the plain lattice is curled
up to a torus.

We here show that although the topological order of decoupled
Ising chains with a periodic boundary condition is trivial, when
the in-chain coupling constants are special chosen with respect to
the chemical potential, there are low-lying non-local excitations
in an open boundary condition which may obey anyonic statistics.
We will see that the low-lying excitations of this model are
classified by several kinds of closed subsets. One kind of them is
local, including a single hole, a double occupant and a spin-flip.
Other two kinds consist of a local fermion and two string-like
non-local excitations. The fusion rules and exchange phase factor
in later two are exactly the same as those of the excitations in
the toric code model. The statistics of the non-local excitations
themselves are bosonic while it is mutual semionic. We find that
the ground state in this model is stable for a dipole-dipole long
range interactions if inter-chain couplings are negligible. Thus,
this system can be realized in cold dipolar Fermi atoms, e.g.,
rare-earth atoms of Ytterbium \cite{yb}, or the cold fermionic
heteronuclear molecules like $^{40}$K$^{87}$Rb \cite{jin1}, in
optical lattice. With respect to the ground state degeneracy, any
subset of the excitations is accompanied by many energetically
degenerate subsets. We discuss how to peel off a subset from other
degenerate subsets and manipulate anyons by means of the
established cold atom and molecular techniques.

\noindent{\it Two-component fermion model in square lattice } We
consider a simple Hamiltonian for two-component fermions in a
 square lattice (Fig. \ref{fig:Fig.1})
\begin{eqnarray}
H=-\sum_{\langle ij\rangle_{hd},s}J_s(2n_{s,i}-1)(2n_{s,j}-1)
+U\sum_i (2n_{\uparrow,i}-1)(2n_{\downarrow,i}-1),
\end{eqnarray}
where $n_{s,i}=c^\dag_{s,i}c_{s,i}$ with $c_{s,i}$ being
annihilation operators of (pseudo)spin-$s$ fermions. The symbol
$\langle ij\rangle_{hd}$ means the sum is over nearest neighbors
along the horizontal diagonals of squares. This is a set of
decoupled Ising chains along the horizontal diagonals of squares.
We restrict on the nearest neighbor attractive interaction while
on-site is repulsive, i.e., $J_s>0$ and $U>0$. In the case, the
ground states of this Hamiltonian are $2^n$-fold degenerate, i.e.,
every individual chain is ferromagnetic, i.e., for a set of spins
$\{s_1,\cdots,s_a,\cdots,s_n\}$,
\begin{eqnarray}
|G_{\{s\}}\rangle=\prod_{a=1}^n|G_{s_a}\rangle=\prod_{a,i_a}c^\dag_{s_a,i_a}|0\rangle,
\end{eqnarray}
where $n$ is the number of chains, $i_a$ is the site index in the
$a$-th chain. Restricted to the open boundary condition, the
low-lying excitations above a given ground state
$|G_{\{s\}}\rangle$, for a given site $i_a$, reads (See Fig. 1 for
a given ground state $|G_\uparrow\rangle$ which will be defined
later.)
\begin{eqnarray}
{\cal
H}_{i_a}|G_{\{s\}}\rangle&=&(c^\dag_{s_a,i_a}+c_{s_a,i_a})|G_{\{s\}}\rangle=c_{s_a,i_a}|G_{\{s\}}\rangle,\nonumber\\
{\cal D}_{i_a}|G_{\{s\}}\rangle&=&(c^\dag_{\bar s_a,i_a}+c_{\bar
s_a,i_a})|G_{\{s\}}\rangle
=c^\dag_{\bar s_a,i_a}|G_{\{s\}}\rangle,\nonumber\\
{\cal F}_{i_a}|G_{\{s\}}\rangle&=&i{\cal H}_{i_a}{\cal
D}_{i_a}|G_{\{s\}}\rangle=ic_{s_a,i_a}c^\dag_{\bar
s_a,i_a}|G_{\{s\}}\rangle,\nonumber\\
{\cal W}_P|G_{\{s\}}\rangle&=&\prod_{i'_b\leq i_a} {\cal
F}_{i'_b}|G_{\{s\}}\rangle=\prod_{i'_b\leq i_a}
ic_{s_b,i'_b}c^\dag_{\bar s_b,i'_b}|G_{\{s\}}\rangle,\nonumber\\
{\cal W}_{P,P'}^h|G_{\{s\}}\rangle&=&\prod_{i'_b<i_a} {\cal
F}_{i'_b}{\cal
H}_{i_a}|G_{\{s\}}\rangle=\prod_{i'_b<i_a}ic_{s_b,i'_b}c^\dag_{\bar
s_b,i'_b}c_{
s_a,i_a}|G_{\{s\}}\rangle,\label{exc}\\
 {\cal W}_{P,P'}^d|G_{\{s\}}\rangle&=&\prod_{i'_b<i_a} {\cal
F}_{i'_b}{\cal
D}_{i_a}|G_{\{s\}}\rangle=\prod_{i'_b<i_a}ic_{s_b,i'_b}c^\dag_{\bar
s_b,i'_b}c^\dag_{\bar s_a,i_a}|G_{\{s\}}\rangle,\nonumber
\end{eqnarray}
where $P$ and $P'$ denote two plaquettes on the right and left of
$i_a$, respectively. $\bar s=\downarrow(\uparrow)$ if
$s=\uparrow(\downarrow)$. The order of sites is defined by
$i'_b<i_a$ if $i'_b$ is on the left hand of $i_a$ for $b=a$ or
$i'_b$ is in a chain lower than the chain with $i_a$. ${\cal
H,D,F}$ create a hole, a double occupant, and a spin-flip. ${\cal
W}$, ${\cal W}^d$ and ${\cal W}^h$ create a half-infinite string
of spin-flips, a spin-flip string with a double occupant and a
spin-flip string with a hole, respectively, since the spins of
fermions at sites
 $i'_b<i_a$ are flipped from their ground state configuration while those
at $j_c>i_a$ keep in their ground state configuration. The
excitation energies of these local and non-local excitations
 in turn are $4J_{s_a}+2U$,
$4J_{\bar s_a}+2U$, $4J_{\uparrow}+4J_{\downarrow}$,
$2J_{\uparrow}+2J_{\downarrow}$,
$2J_{\uparrow}+2J_{\downarrow}+2U$ and
$2J_{\uparrow}+2J_{\downarrow} +2U$, respectively. The finite
energies of the string-like excitations mean they are deconfined.
Note that the ground state may be instable against an on-site
energy $V_B\sum_i(2n_{\uparrow(\downarrow),i}-1)$ if $V_B>0$ and
if $V_B<0$, a pair of string-like excitations may be linearly
confined and the statistics of the string-like excitations becomes
ill-defined. These excitations are highly degenerate due to the
degeneracy of the ground states. For example, if
$\{s_1,\cdots,s_a\cdots,s_n\}\to\{s_1,\cdots,\bar
s_a\cdots,s_n\}$, ${\cal H}\leftrightarrow {\cal D}$ and
$(\leq,<)\to (\geq,>)$ in (\ref{exc}). One can also flip spin in
other chains to get a new degenerate excitation.

\noindent{\it Fusion rules} Note that ${\cal O}^2=I$ for ${\cal
O}={\cal H,D,F,W,W}^d,$ ${\cal W}^h$. The fusion rules of these
excitations are given in Table 1. The fusion rules of the closed
subset $\{I, {\cal D}_{i_a}, {\cal W}_P,{\cal W}_{P,P'}^h\}$ (or
$\{I, {\cal H}_{i_a}, {\cal W}_P,{\cal W}_{P,P'}^d\}$ ) are
exactly the same as the fusion rules in Kitaev toric code model if
we identify ${\cal D}_{i_a}, {\cal W}_P$, and ${\cal W}_{P,P'}^h$
as $\psi, e$ and $m$ \cite{kitaev,kitaev1}. The subset $\{I, {\cal
F}_{i_a}, {\cal W}_{P,P'}^h,{\cal W}_{P,P'}^d\}$ has similar
fusion rules but ${\cal F}$ is  bosonic. $\{I,{\cal H}_{i_a},{\cal
D}_{i_a}, {\cal F}_{i_a}\}$ is also a closed subset and has
similar fusion rules but with two fermions (${\cal H,D}$) and one
boson (${\cal F}$).

{\small

\def\arraystretch{0}\begin{tabular}{|c|c|c|c|c|c|c|c|c|}
  \hline &${\cal H}_{i_a}$ &${\cal D}_{i_a}$&${\cal F}_{i_a}$&${\cal
W}_P$& ${\cal W}_{P,P'}^h$
&${\cal W}_{P,P'}^d$ &${\cal W}_{P'}$ \\
   \hline  ${\cal H}_{i_a}$ &$I$ &${\cal F}_{i_a}$ &${\cal
D}_{i_a}$ &${\cal W}^d_{P,P'}$ & ${\cal W}_{P'}$
 &${\cal W}_{P}$ &${\cal W}_{P,P'}^h$ \\
    \hline  ${\cal D}_{i_a}$ &${\cal F}_{i_a}$  &$I$ &${\cal H}_{i_a}$
&${\cal W}_{P,P'}^h$ &${\cal W}_P$
  & ${\cal W}_{P'}~$
 &${\cal W}_{P,P'}^d$ \\
    \hline  ${\cal F}_{i_a}$ &${\cal D}_{i_a}$ &${\cal H}_{i_a}$ &$I$
&${\cal W}_{P'}$  & ${\cal W}_{P,P'}^d$
 &${\cal W}_{P,P'}^h$ &${\cal W}_{P}$~ \\
     \hline ${\cal W}_{P}$ &${\cal W}_{P,P'}^d$
&${\cal W}_{P,P'}^h$ &${\cal W}_{P'}$ &$I$
 & ${\cal D}_{i_a}$
 &${\cal H}_{i_a}$ &${\cal F}_{i_a}$  \\
   \hline $~{\cal W}_{P,P'}^h$ &${\cal W}_{P'}$
&${\cal W}_{P}$ &${\cal W}_{P,P'}^d$ &${\cal D}_{i_a}$
 & $I$
 &${\cal F}_{i_a}$ &${\cal H}_{i_a}$  \\
   \hline $~{\cal W}_{P,P'}^d$ &${\cal W}_{P}$
&${\cal W}_{P'}$ &${\cal W}_{P,P'}^h$ &${\cal H}_{i_a}$
 & ${\cal F}_{i_a}$
 &$I$ &${\cal D}_{i_a}$  \\
   \hline ${\cal W}_{P'}$
&${\cal W}_{P,P'}^h$&${\cal W}_{P,P'}^d$ &${\cal W}_{P}$ &${\cal
F}_{i_a}$
 & ${\cal H}_{i_a}$
 &${\cal D}_{i_a}$ &$I$ \\
\hline
\end{tabular}

\vspace{2mm}

 \centerline{Table 1: Fusion rules of excitations.}}

\vspace{1mm}

\noindent{\it Discrete gauge symmetry} The conserved quantities
are simply $Q_{s,i}=2n_{s,i}-1$, (in fact, they are $n_{s,i}$ ),
which have eigen values $\pm 1$ at each site. This generates
$Z_2\times Z_2$ gauge symmetry. For a plaquette $P$, one can label
the plaquette by $G^{s_a}_P=(2n_{i_a,s_a}-1)(2n_{j_a,s_a}-1)$ for
a pair of nearest neighbors $(i_a,j_a)$ in the $a$-th chain, which
also have eigen values $\pm 1$. Obviously, the ground state is of
all $G_P^s=1$. The excitations are also of all $G^s_{P"}=1$ except
for the right plaquette $P$ and left plaquette $P'$ of $i_a$ where
$(G^{s_a}_P,G^{\bar s_a}_P,G^{s_a}_{P'},G^{\bar
s_a}_{P'})=(-1,1,-1,1)$ for ${\cal H}_{i_a}$, $(1,-1,1,-1)$ for
${\cal D}_{i_a}$, $(-1,-1,-1,-1)$ for ${\cal F}_{i_a}$,
$(1,1,-1,-1)$ for ${\cal W}_P$, $(\pm 1,\mp 1,\mp 1,\pm 1)$ for
${\cal W}_{P,P'}^h$ and $(\mp 1,\pm 1,\pm 1,\mp 1)$ for ${\cal
W}_{P,P'}^d$. To distinguish latter two, one uses $(Q_{s_a,i_a},
Q_{\bar s_a,i_a})=(-1,-1)$ for ${\cal W}_{P,P'}^h$ and $(1,1)$ for
${\cal W}_{P,P'}^d$. In this sense, these string-like excitations
are also called $Z_2\times Z_2$ vortices.

\noindent{\it Boundary conditions} A string-like excitation may be
thought as a spin flipping domain wall, a topological defect (See
Fig. 1, late four). The open boundary condition allows odd number
of domain walls while the periodic boundary condition restricts
the wall number to even in each chain. If we curl up the plain
lattice to a torus, the topological order is completely trivial
and the system is identical to a one-dimensional one. In the open
boundary condition,  two walls as ends of a string, e.g, $
\cdots\uparrow\uparrow\circ\downarrow\downarrow\cdots$ and $
\cdots\downarrow\downarrow\uparrow\uparrow\cdots$, may locate at
different chains
  so that one can circle around another. Therefore,
their (mutual) exchange statistics can be well defined.

\begin{figure}[htb]
\includegraphics[width=6cm]{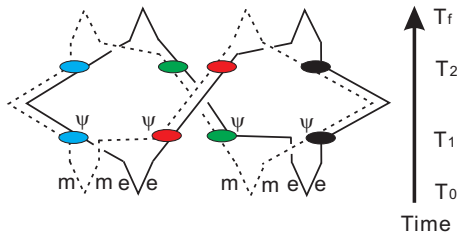}
 \caption{\label{fig:Fig.2} (Color on line) Two pairs of $m$ (dashed lines)
 and two pairs of $e$ (solid lines)
created from the ground state at $T_0$; they fuse into four
fermions $\psi$ (colored ellipses) at $T_1$; then they split back
to four pairs at $T_2$ and annihilate to the ground state at
$T_f$. The green and red fermions exchange, which differ a minus
sign from none of exchange. The arrow indicates the time
direction.}
\end{figure}

\begin{figure}[htb]
\includegraphics[width=3cm]{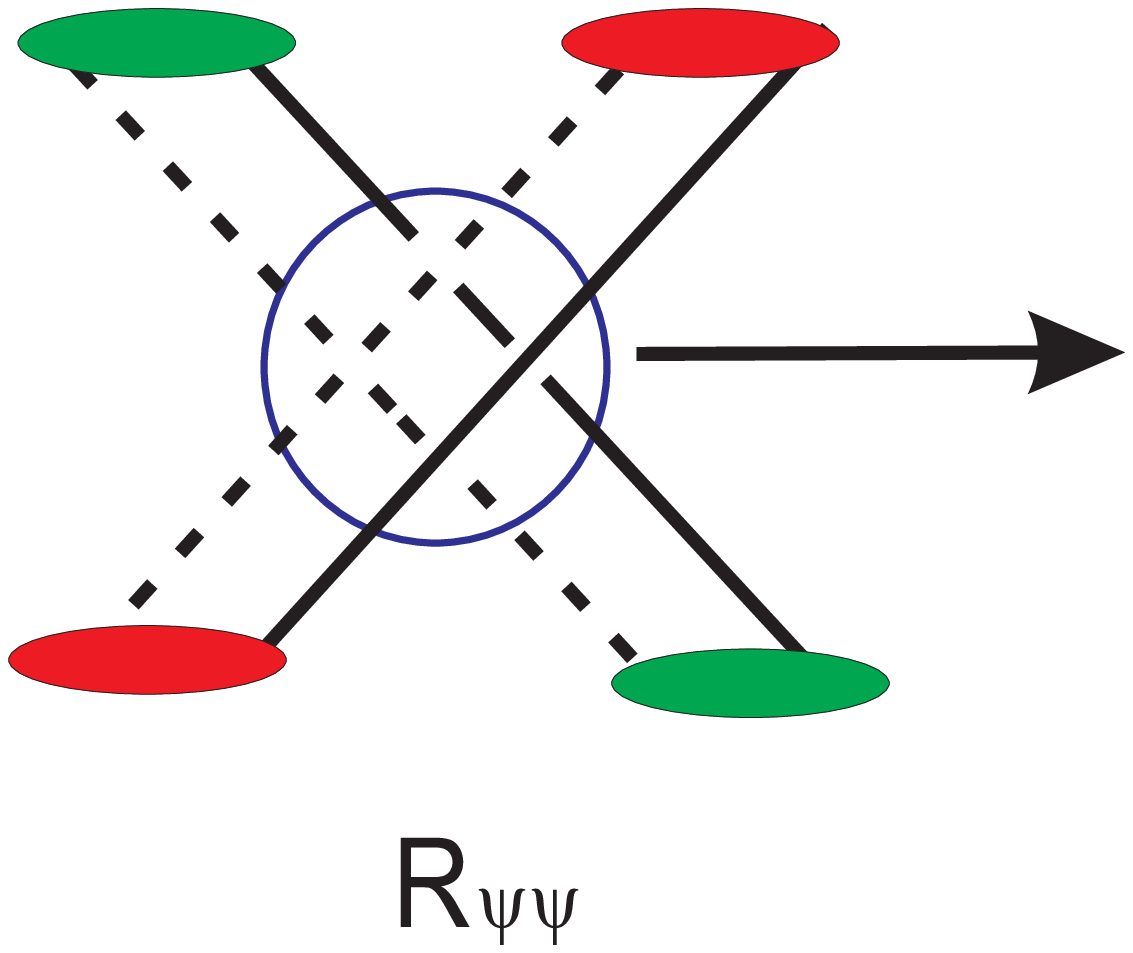}
\includegraphics[width=3cm]{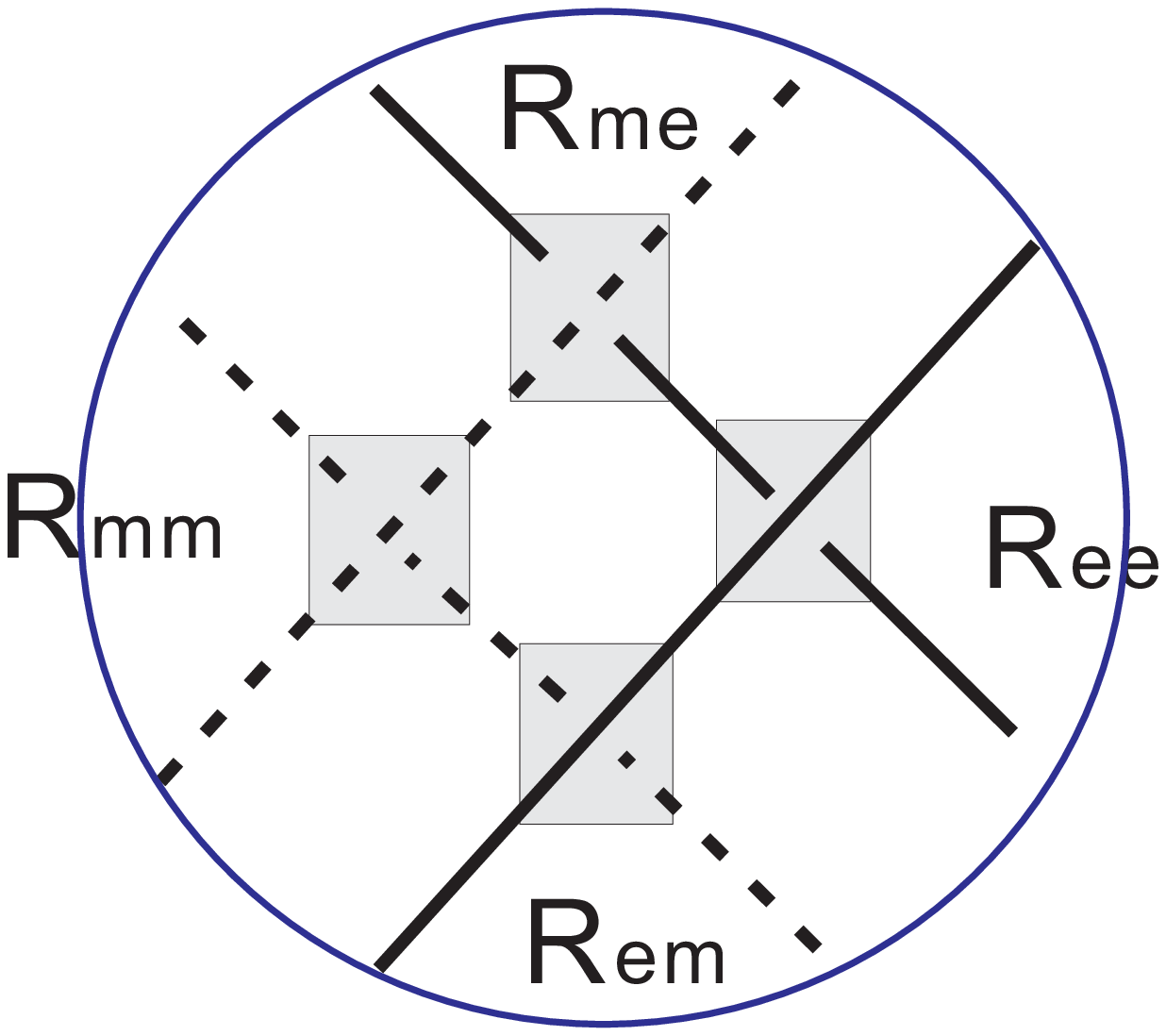}
 \caption{\label{fig:Fig.3} (Color on line) Divide
 the fermion exchange $R_{\psi\psi}$
 into exchanges between the same non-local excitations,
 $R_{ee}$ and $R_{mm}$,
 and double exchange $e$ and $m$,
 $R_{me}R_{em}$, which is equal to moving $e$ (or $m$) around $m$ (or $e$).
}
\end{figure}

\noindent{\it Statistics in a given subset} We now study the
statistics of the non-local excitations within a subset. The local
excitations ${\cal H, D}$ are the fermionic while ${\cal F}$ is
bosonic. For a string-like excitation, the site of its domain
wall, $i_a$,  may be used to label its 'position', e.g., ${\cal
W}_P={\cal W}_{i_a}$, etc. ${\cal W}$ is bosonic because it is a
string of ${\cal F}$. ${\cal W}^{d,h}$ themselves are bosonic,
e.g.,
\begin{eqnarray}
{\cal W}_1^h{\cal W}_2^h=({\cal H}_1)({\cal H}_1{\cal D}_1{\cal
H}_2)=({\cal H}_1{\cal D}_1{\cal H}_2)({\cal H}_1)={\cal
W}_2^h{\cal W}_1^h.\nonumber
\end{eqnarray}
Since ${\cal W}$ and ${\cal W}^{d,h}$ are distinguishable, the
exchange between them does not make sense. However, because ${\cal
WW}^{h,d}$ fuses to a fermion while themselves are bosons, when
${\cal W}$ circles around ${\cal W}^{h,d}$ or vice verse, a minus
sign is acquired. In general, this fact can be proved by applying
the consistent conditions, i.e., the pentagon and hexagon
equations \cite{phe}. For the present case, one can take Kitaev's
graphical proof \cite{kitaev1}. For simplicity, denote $\{I,{\cal
H,W,W}^d\}(~{\rm or}~\{I,{\cal D,W,W}^h\})=\{I,\psi,e,m\}$. Since
$e^2=m^2=I$, we can create two pairs of $e$ and two pairs of $m$
from a given ground state at an initial time $T_0$ (See Fig.
\ref{fig:Fig.2}). These excitations move along the pathes as shown
in Fig. \ref{fig:Fig.2}. At certain time $T_1$, they fuse to four
fermions $\psi$. As time flies, while blue and black fermions stay
alone, green and red fermions exchange their positions. Finally,
at $T_2$, the fermions split to $m$ and $e$ pairs which, at the
end ($T_f$), fuse back to the ground state. As two fermions
exchange, this process contributes a minus sign $R_{\psi\psi}=-1$
to the ground state comparing to a process without fermion
exchange. Now, examine this process in non-local excitation
exchanges. As shown in Fig. \ref{fig:Fig.3}, this fermion exchange
is corresponding to four exchanges: $R_{em}, R_{ee},R_{mm}$ and
$R_{me}$. That is,
\begin{eqnarray}
 R_{me}R_{ee}R_{mm}R_{em}=R_{me}R_{em}=R_{\psi\psi}=-1,
 \end{eqnarray}
 since $R_{ee}=R_{mm}=1$ as $e$ and $m$ themselves are bosonic.
 The minus sign when $e$ and $m$ doubly exchange, or equivalently,
 $e$ encircles $m$, $R_{me}R_{em}=-1$,
 proves the mutual statistics between $e$ and $m$ is
 semionic.
Note that for the subset $\{I,{\cal F,W}^d,{\cal W}^h\}$, since
 ${\cal F}={\cal W}^d{\cal W}^h$ is bosonic,  $R_{{\cal W}^d{\cal
 W}^h}R_{{\cal W}^h{\cal
 W}^d}(=R_{\cal FF}=1)$ is trivial.


\noindent{\it Peeling off a subset from degeneracy} So far, we
only say string-like excitations obey mutual semionic statistics
but not call them anyons or semions. The reason for this is there
are many degenerate non-local states which are not in the same
subset. For example, flipping any chain's spin for ${\cal W_P}$
results in a degenerate excitation with ${\cal W_P}$ but already
out of the subset of ${\cal W_P}$.  In this sense, these
excitations can not be identified as individual quasiparticles.

To peel off a designed subset, we need to set barriers between
individual degenerate ground states without changing their
energies. To control the electron spin of each individual chain is
not easy. However, it becomes possible in a cold atom (molecule)
system because the 'spin' we are studying actually labels the
different hyperfine states of atoms in the cold atom context .
Once an atom is in a given hyperfine state, local fluctuations
from the environment can not switch it to others. Therefore, we
can peel off a given string-like excitation from others by
preparing the ground state. For example, we can apply a magnetic
field so that only the atoms with a given hyperfine state are
loaded into the lattice and then turn off the magnetic field after
the system is stable at the ground state. A global ferromagnetic
ground state
$|G_\uparrow\rangle=|G_{\{\uparrow,\uparrow,\cdots,\uparrow\}}\rangle$
is prepared. All excitations in (\ref{exc}) then can be prepared
by creating, annihilating the fermions or changing fermions from
$\uparrow$  to $\downarrow$ hyperfine states by means of recently
developed stimulated Raman spectroscopy or photoemission
spectroscopy technique\cite{dao}. Here annihilating and creating a
fermion do not means removing fermions from or reloading them into
lattice sites. It can be turned into other hyperfine states which
are almost not coupled to the 'spin' $\uparrow$ and $\downarrow$
hyperfine states or reverse. The non-local excitations may be
controllably prepared. We may merely prepare excitations in a
given subset and they are barricaded from their degenerate states.
The non-local excitations obeying mutual simonic statistics in
this subset are now identified as {\it mutual semions}.

Recently, a realization of this model in superconducting circuits
has been proposed \cite{xue}. It is also a possible way to peel
off a semionic subset.

\noindent{\it Cold fermions with dipole-dipole interaction} For
cold fermions in optical lattice, off-site interaction between the
cold atoms can be induced by their diploe-dipole interaction.
Recently, the degenerate Fermi gases of rare-earth atoms of
Ytterbium (Yb) have been obtained \cite{yb}. They are possible
candidate to be a practical system of our model because the
fermionic isotopes $^{171}$Yb and $^{173}$Yb are stable in nature
and their metastable state $^3$P$_2$ has a large magnetic dipole
moment $3\mu_B$.  Deeply bound cold fermionic heteronuclear
molecules have much larger dipole moment, e.g., the electric
dipole moment of $^{40}$K$^{87}$Rb in its absolute bound ground
state is 0.3 $ea_B$ \cite{jin1}. Load the fermions in an optical
lattice and polarize all dipoles along the horizontal diagonal of
squares by using an external field. The interaction potential is
$V_d({\bf r},{\bf r}')=d^2\frac{1-3\cos^2\Theta}{R^3}$ where
$\Theta$ is the angle between ${\bf R}={\bf r}-{\bf r}'$ and ${\bf
d}$ (the diploe moment of an atom). The interactions along the
diagonal become attractive. The repulsive interaction is
restricted in the region with $\Theta>\Theta_c$ for
$\cos^2\Theta_c=1/3$. The interacting Hamiltonian can be written
as
\begin{eqnarray}
V=-\sum_{\langle
ij\rangle_{hd},s}|V_{ij,s}|n_{s,i}n_{s,j}-\sum_{\langle\langle
ij\rangle\rangle_{hd},s}|V_{ij}|n_{s,i}n_{s,j}
-\sum_{ij,0<\Theta<\Theta_c,s}|V_{ij}|
n_{s,i}n_{s,j}+\sum_{ij,\Theta>\Theta_c,s}V_{ij} n_{s,i}n_{s,j},
\end{eqnarray}
where $\langle\langle ij\rangle\rangle_{hd}$ denotes the sum along
the horizontal diagonal other than the nearest neighbors.  It is
very easy to stabilize the ground state because one may increase
the distance between the horizontal chains or adjust the optical
lattice potentials so that the inter-chain couplings become weak.

Strictly speaking, the anyons emerging from these dipolar particle
systems are logarithmic confinement in thermodynamic limit,
 i.e., $E_{\rm
pair}-E_g\propto \ln L$ as $L\to \infty$ for $L$ the distance
between a pair of the non-local excitations. This weak divergence
in a practical optical lattice may be abided, e.g., if the short
range model we proposed has $E_{\rm pair}-E_g\sim 1$, this
logarithmic excitation energy is $4.5$ for $L=50$. Even $L=1000$,
this energy only increases about 15 times.

\noindent{\it Conclusions} In conclusions, we have proved that it
is possible to find a non-trivial mutual anyonic statistics in a
fermionic system with conventional two-body interaction in open
boundary condition. How to peel up a semionic quasiparticle from
many degenerate states and manipulate them were discussed.

\noindent{\it Acknowledgements} The authors thank J. P. Hu, X. L.
Qi, X. R. Wang, Z. D. Wang, C. J. Wu, K. Yang, S. Yi, G. M. Zhang
and S. C. Zhang for the useful discussions. One of authors (Y. Y.)
thanks Professor Ruibao Tao and Physics Department, Fudan
University for the warm hospitality. This work was supported in
part by the NNSF of China, the national program for basic research
of MOST of China and a fund from CAS.

\end{document}